\documentclass[preprint2]{aastex}

\newcommand{\myemail}{rmason@gemini.edu}

\newcommand{\be}{\begin{equation}}   
\newcommand{\ee}{\end{equation}}     
\def\deg{\hbox{$^{\:\circ}$}}

\shorttitle{L-band spectropolarimetry of NGC1068} 

\shortauthors{Mason et al.}

\begin{document}
\title{Spectropolarimetry of the 3.4~$\mu$m absorption feature in NGC~1068}

\author{R. E. Mason} \affil{Gemini Observatory Northern Operations Centre, 670 N. A'ohoku Place, Hilo, HI 96720, USA}
\email{\myemail}

\author{G. S. Wright} 
\affil{UK Astronomy Technology
center, Royal Observatory Edinburgh, Blackford Hill, Edinburgh EH9 3HJ, UK}
\email{gsw@roe.ac.uk}

\author{A. Adamson}
\affil{Joint Astronomy Centre, 660 North A`ohoku Place, Hilo, HI 96720, USA}
\email{a.adamson@jach.hawaii.edu}

\and

\author{Y. Pendleton}
\affil{NASA Ames Research Center, Mail Stop 245-3, Moffet Field CA 94035, USA}
\email{ypendleton@mail.arc.nasa.gov}

\begin{abstract}

In order to test the silicate-core/organic-mantle model of galactic interstellar dust, we have performed spectropolarimetry of the 3.4 $\mu$m C--H bond stretch that is characteristic of aliphatic hydrocarbons, using the nucleus of the Seyfert 2 galaxy, NGC1068, as a bright, dusty background source. Polarization calculations show that, if the grains in NGC1068 had the properties assigned by the core-mantle model to dust in the galactic diffuse ISM, they would cause a detectable rise in polarization over the 3.4 $\mu$m feature. No such increase is observed.  We discuss modifications to the basic core-mantle model, such as changes in grain size or the existence of additional non-hydrocarbon aligned grain populations, which could better fit the observational evidence.  However, we emphasize that the absence of polarization over the 3.4 $\mu$m band in NGC1068 --- and, indeed, in every line of sight examined to date --- can be readily explained by a population of small, unaligned carbonaceous grains with no physical connection to the silicates.

\end{abstract}


\section{Introduction}

\label{sect:intro}

In our galaxy and no doubt others, stars and planets form from components available in the local 
interstellar medium. The organic compounds observed in
interstellar space may therefore be the first step towards the complex materials
that help make planets habitable. A greater understanding of the origin
and evolution of the organic materials in the ISM,  both in our galaxy and in others, is thus
of great interest.

Particularly puzzling are the aliphatic hydrocarbons, whose presence
in the diffuse ISM (DISM) of the Milky Way has been established through observations of
C-H bond stretches near 3.4 $\mu$m, seen in absorption throughout the
DISM of our own galaxy and others
\citep{S91,P94,S95,Im00b,Rawlings03,Ris,Dogtanian,Mason04}. Two very different models exist to
explain the production of this material: photoprocessing of the icy
dust grain mantles found in dense molecular clouds \citep[e.g.][]{Li97},
and hydrogenation in the DISM of small, amorphous 
carbon particles of the kind produced
in outflows from evolved stars \citep[e.g.][]{Me02}. As the former
process would result in the 3.4 $\mu$m band of DISM hydrocarbons
arising in organic mantles on silicate grain cores, it is commonly
known as the ``core-mantle'' dust model.

Both of these models are to some extent consistent with the
observational data. For instance, laboratory experiments mimicking both
dust formation pathways have succeeded in producing materials with
3.4 $\mu$m bands which are a good match to the astronomical
feature \citep{Greenberg95,Me99,Me02}, although the critically diagnostic 5-10 $\mu$m region reveals a much closer match between observations and hydrogenated amorphous carbon (HAC) materials \citep{Chiar00,PandA}.
The presence of the 3.4 $\mu$m band in a
protoplanetary nebula \citep{Chiar98} shows that 
hydrocarbon solids can be produced in circumstellar regions, while the attribution of the 6 $\mu$m excess absorption in protostars to organic refractory material \citep{GW02} would, if confirmed, indicate that ice processing can also produce some fraction of the hydrocarbons. The organic OCN$^{-}$ band at 4.62~$\mu$m  provides evidence of ice processing in such environments \citep{Demyk,P99}.

Spectropolarimetry provides opportunities for clearly discriminating
between these scenarios. In the core-mantle model, the silicate
core/organic mantle grains are responsible for the bulk of the visual and IR
extinction and all of the polarization in the DISM \citep{Li97}. As the
polarization is caused by dichroic absorption, we would therefore
expect to observe polarization of the continuum, with an increase in
polarization over absorption features associated with the 
grains \citep[e.g.][]{H88,PTW,Smith00}. Specifically,  increases in polarization must occur
simultaneously over the 3.4 $\mu$m ``mantle'' feature, and the
9.7 $\mu$m absorption characteristic of the silicate core, with roughly
comparable efficiency \citep{LG02}. This was first investigated by
\citet{A99}, who compared the polarization over the 3.4 $\mu$m feature
towards the Galactic center source, IRS 7, with that of the 9.7 $\mu$m
band towards the nearby Galactic center source, IRS 3. The increase in
polarization over the 3.4 $\mu$m absorption was found to be $\la$0.08\%,
much less than the 0.4\% expected on the basis of the silicate feature
polarization, in apparent conflict with the core-mantle
model. 

However, as pointed out by \citet{LG02}, the silicate feature
polarization used in that work, while arising toward an object at a projected distance of
about 0.25pc from that towards which the 3.4 $\mu$m polarization was
measured \citep{Geballe89}, does not refer to exactly the same line of
sight, admitting the possibility that the silicate feature in the diffuse ISM towards
IRS 7 may be less polarized than expected.
Studies of additional lines of sight (at lower signal-to-noise ratio and/or spectral resolution)
while reaching similar conclusions to those of \citet{A99} , suffer from similar limitations \citep{Nagata94,Ishii}. More recently, \citet{Chiar06} measured the polarization of both the hydrocarbon and silicate features towards the Quintuplet Cluster in the Galactic Center, finding the hydrocarbons to be significantly less polarized than would be expected if they existed as mantles on the polarized silicate grains.

To provide another data point against which the predictions of the core-mantle grain model can be compared, and to extend this study to a new and different environment, we have used the bright nucleus of the Seyfert 2 galaxy NGC1068 as a source against which to measure the polarization
of the 3.4 $\mu$m band and the continuum around it.  Rather than the 10~$\mu$m silicate feature, we have used the degree of continuum polarization to calculate the enhancement in polarization that would be expected over the 3.4~$\mu$m feature from a screen of elongated,
coated silicate grains which polarize by the selective absorption of one plane of the incident light. 
In \S\ref{los}, we describe the line of sight towards the nucleus of
NGC1068 and its suitability for this work. The spectropolarimetric
observations and their treatment are outlined in \S\ref{obsdr}. The calculations that we have carried out are discussed in \S\ref{calcs}, and
the results presented in \S\ref{results}. This work is then summarized in
\S\ref{conc}.

\section{Dust and dichroic polarization in NGC~1068}
\label{los}

As the archetypal Seyfert 2 galaxy, NGC~1068
 (d=16 Mpc; $\rm H_{0}=70 km \ s^{-1} \ Mpc^{-1}$; 1\arcsec\ = 72 pc) 
has been the center of much
attention. The unified model of active galactic nuclei (AGN) implies that the galaxy
harbors a Seyfert 1 nucleus,  obscured from our point of view
by a torus of dust and molecular gas, and 10~$\mu$m
interferometry \citep{Jaffe} as well as the detection of broad emission lines in polarized light \citep{AM85} have provided strong evidence for the existence of
such a torus.  Copious other mid-IR and X-ray data also show that the nucleus of NGC~1068 is obscured by large amounts of  warm dust \citep[e.g.][]{Bock2,Matt00,Tomo,Alloin,Mason06}.  While the classification of NGC~1068 as a type 2 object suggests that the obscuring torus is oriented roughly edge-on to our line of sight, the disk of the galaxy has quite a low inclination \citep[$i\approx 29\deg$;][]{GG02}. This means
that the line of sight to the center of NGC~1068 samples dust local to the active nucleus,
with a negligible contribution from dust in the disk of the galaxy.

The 3.4 $\mu$m absorption band has been observed in NGC~1068
\citep{B94,Wright96,Imanishi97,MB03,Mason04}, as has silicate absorption at 9.7 $\mu$m
\citep{KGW,Roche84,Jaffe,Mason06,Rhee06}. The 3.4 $\mu$m band has a very similar profile to that observed in Galactic lines of
sight; although the band carrier exists in the central region of an
active galaxy, it apparently undergoes little extra processing
compared to the Galactic DISM \citep{Dogtanian,Mason04}. NGC~1068 has a nuclear dust spiral which appears physically connected to the galaxy disk \citep{Pogge}, so it is quite possible that
the hydrocarbons in the nucleus of NGC~1068 formed in the disk of the galaxy and then migrated to the
center.
We therefore proceed on the assumption that the hydrocarbon-containing
dust in the nucleus of NGC~1068 initially formed  by one of the mechanisms that has been suggested to be
responsible for Galactic aliphatic hydrocarbons.

The nucleus of NGC~1068 is known to be polarized from the UV
to the mid-IR, and the mechanisms responsible for this polarization have been the subject of
intense study. \citet{AM85} were among the first to observe the UV and
visible polarization of NGC~1068, and they interpreted the high,
wavelength-independent polarization as being due to scattering off
very small particles, probably free electrons. \citet{Bailey} later
discussed the polarization properties of NGC~1068 from the UV out to
10$\mu$m, and found a twist in position angle from the optical into
the near-IR. They interpreted this as dust or electron scattering
being replaced by dichroic absorption, the preferential absorption of one component of the electric vector, as the most important polarizing
mechanism. Their data also show a change of $\sim 70^{\circ}$
in position angle between the L and M bands, consistent with dichroic emission
from aligned dust grains becoming the dominant polarizing mechanism at longer wavelengths.

Further evidence for the importance of dichroic absorption as a
polarizing mechanism in the IR comes from imaging polarimetry. After analyzing the deviations from the
centrosymmetric patterns expected from scattering alone,
\citet{Lumsden99} concluded that their data require another mechanism of
constant position angle contributing to the polarization. The effect
of this mechanism grows with increasing wavelength, as would be
expected if absorption were beginning to dominate over scattering, and was
estimated to account for $>90$\% of the K band
polarization. Lumsden at el. were successful in fitting their
observed JHK polarized flux points with greybody emission from hot
($T\sim1200$K) dust, reddened by a $\lambda^{-1.75}$ extinction curve
and scaled by a Serkowski law appropriate for moderately extinguished
Galactic sources.  Furthermore, \citet{Packham97} interpreted the different aperture dependence of the J-, H- and K- band polarization as consistent with a dichroic contribution in the H- and K-bands. With higher-resolution HST imaging polarimetry,
\citet{Simpson02} also found that the K-band polarization has a
contribution from dichroism, but they were able to set tighter limits
on the location of the dichroic component, to within 1\arcsec~of the
nucleus.

This agrees with the more detailed modeling of \citet{Young95}, who
find that neither electron nor dust scattering can account for the
near-IR polarization of NGC~1068. However, if a dichroic component is
added, the near-IR polarized flux spectrum can be reproduced. In this
model there is still some contribution from electron scattering to the
near-IR flux, but this decreases with wavelength relative to the
dichroic component. By the K band, electron scattering contributes
perhaps 10\% of the total flux, with dust making up almost all of the
remainder. At L, the scattering contribution in their model would be
negligible.

\citet{Watanabe} are also successful in modeling the near-IR
polarization of NGC~1068 with polarization from aligned dust grains. In
addition, they raise the possibility that the IR polarization could be
caused by scattering off large grains in the torus, and point out that
neither the 70\deg change in position angle nor the absence of a
centrosymmetric scattering pattern at longer wavelengths necessarily rules out
scattering as the polarizing mechanism. This has yet to be tested in any detail, but the implications that it may have for the conclusions of this study are discussed in \S\ref{results}. 

While the near-IR data point to dust absorption being the dominant polarizing mechanism in the L band, spectropolarimetry of the 9.7~$\mu$m silicate feature does not show the pronounced polarization excess that might be expected if emission or absorption from aligned silicates were causing the polarization \citep{Aitken84}. This observation was interpreted as evidence that the mid-IR polarization arises in emission from non-silicate grains or has a nonthermal origin.  However, the lack of polarization in the silicate feature does not necessarily rule out the presence of aligned silicate-core grains in the nucleus of NGC~1068.  In radiative transfer calculations of dichroic polarization from silicate-containing grain mixtures in dusty disks,  \citet{Efstathiou97} and \citet{Aitken02} find numerous configurations in which  the polarization over the feature is in fact quite flat, consistent with the observations of \citet{Aitken84}.  

Although it appears possible for radiative transfer effects to suppress the silicate feature polarization, the shorter-wavelength 3.4~$\mu$m band should be much less affected by such interplay between absorption and emission. The ratio of the optical depths of the 3.4 and 9.7$\mu$m bands in NGC~1068 supports this suggestion. In the five Galactic lines of sight examined by \citet{S95}  there is a fair degree of correlation between the depths of the two features ($\tau_{9.7}/\tau_{3.4}=13-19$, with the higher values towards the Galactic Centre), but in NGC~1068  $\tau_{9.7}/\tau_{3.4}\approx 5$, suggesting that the 3.4~$\mu$m band is indeed less affected by underlying emission than is the silicate feature. This further suggests that treating the dust producing the 3.4~$\mu$m band as a uniform absorbing screen (an assumption implicit in the calculations outlined in \S~\ref{calcs}), while undoubtedly a major simplification of the dust geometry and temperature structure in this AGN,  is still a useful approximation. Overall, imaging polarimetry, spectral
modeling and the near-90\deg~rotation in position angle all imply that
a model of the L-band polarization of NGC~1068 based on selective dust
absorption is a reasonable representation of the true situation.

%££££££££££££££££££££££££££££££££££££££££££££££££££££££££££££££££££££

\section{Observations and Data Reduction}
\label{obsdr}

L-band spectropolarimetry of the nucleus of NGC1068 was obtained on
the nights of 2000 September 19 and 2000 October 6 using the
IRPOL2 spectropolarimetry module and CGS4 on the 3.8m UK Infrared
Telescope on Mauna Kea, Hawaii. The 40 l/mm grating and
0.6\arcsec-wide slit were used, providing R=1360 at 3.4 $\mu$m. The weather conditions were good during
the second night, but some thin cirrus was present on the first\footnote[1]{Spectropolarimetry of the 9.7 $\mu$m silicate feature was later also attempted, but poor weather meant that none of the data obtained were useful.}.

In order to obtain Stokes q and u parameter spectra a frame was taken
with IRPOL2's half-wave plate at each of four positions: 0$^{\circ}$,
45$^{\circ}$, 22.5$^{\circ}$, and 67.5$^{\circ}$. This cycle was
repeated 64 times in all. At each of these waveplate angles, ordinary
and extraordinary beams were extracted from the sky-subtracted frames, stacked together, and the final, total spectra combined using the ``ratio'' method \citep[see e.g.][]{Tin} to produce Stokes
q and u parameter spectra.  This method has the advantage of
minimizing the effects of variations in sky transmission during the
observations.

Prior to calculation of the polarization, further treatment of the raw
q and u spectra was necessary. Firstly, data points between
3.3 and 3.4 $\mu$m (observed) were rejected. This part of the spectrum
contains deep absorptions from the hydrocarbon cement in IRPOL2's
Wollaston prism (at 3.35-3.4 $\mu$m) and from atmospheric methane
(3.31-3.33 $\mu$m), and these lines do not ratio out at all
satisfactorily.

A second effect which must be dealt with is a large-amplitude ripple
in the spectrum which is thought to be caused by multiple reflections
in the waveplate. The ripple was removed using the FFT technique
described by \citet{A99}. Briefly, the ripple causes peaks to appear
around 0.5 and 1.0 times the Nyquist frequency in the Fourier
transform of the spectrum. These peaks were set to zero, then the
transform was inverted.  The instrumental zero point of polarization
was determined and removed from the NGC1068 q and u spectra using observations of the unpolarized star HD18803 \citep{HD18803}, and the resulting polarization spectrum corrected for the imperfect efficiency of the waveplate\footnote[2]{ as given by\\
www.jach.hawaii.edu/UKIRT/instruments/irpol/CGS4/cgs4pol.html\#5}.
The position angle of polarization (PA) was 
calibrated using observations of the young stellar object, AFGL2519,
whose PA at 3-4 $\mu$m has previously been measured by \citet{Hough89}.  An intensity spectrum was also constructed using only a few frames taken near in time to HD18803, which was used as a telluric standard. 

\begin{figure} [htb]
\begin{centering}
\includegraphics[width=80mm,angle=0]{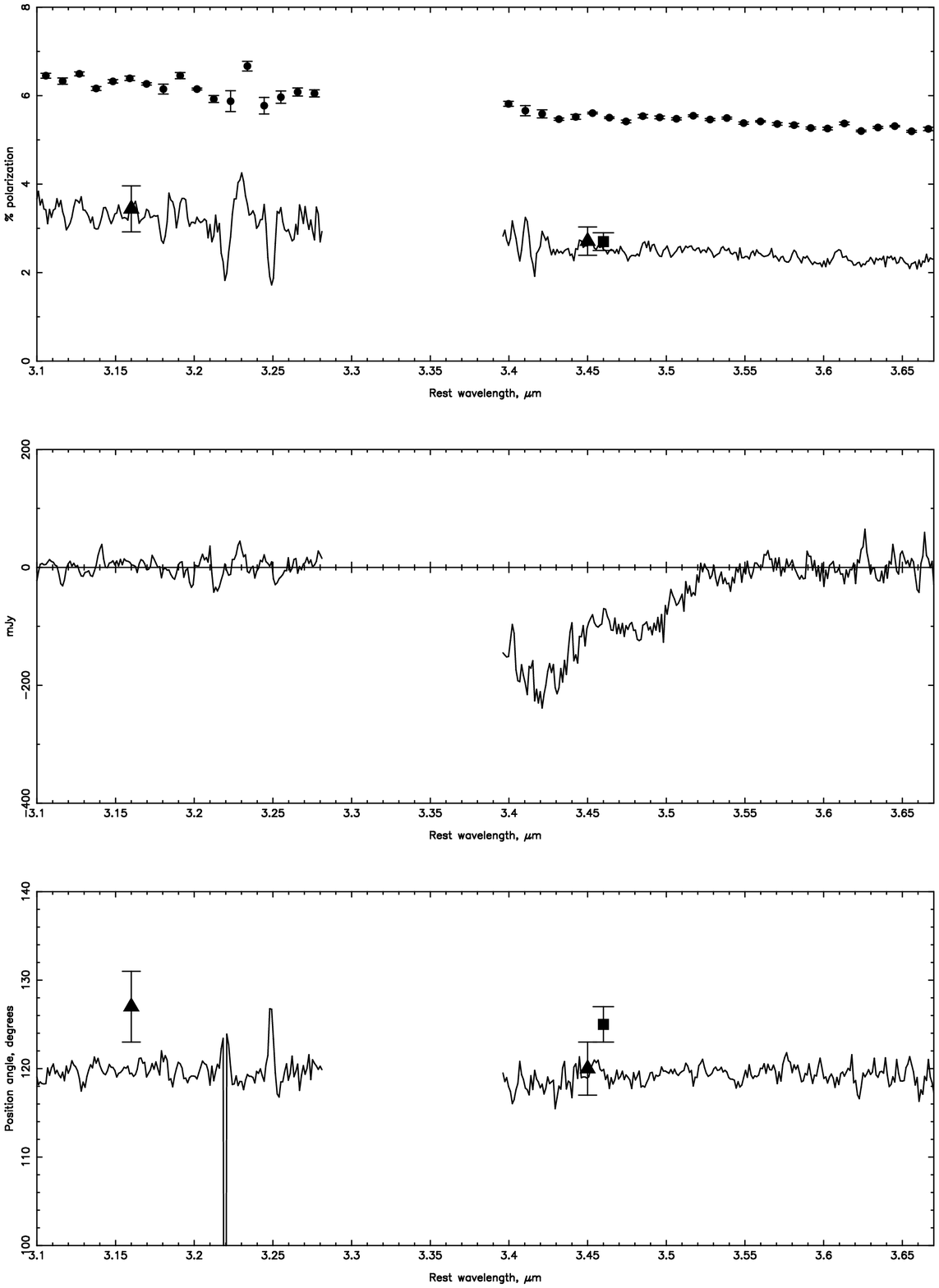}
\caption{\small{Top: L-band polarization spectrum of NGC~1068. The binned spectrum was obtained by calculating the standard error of the q and u spectra in 0.010~$\mu$m bins and is offset
by 3\% for clarity. Also shown are the measurements of \citet[][ triangles]{Bailey} and \citet[][ square, offset by 0.01~$\mu$m in wavelength]{Lebofsky}.} Middle: continuum-subtracted flux spectrum, showing the peak of the 3.4~$\mu$m absorption feature. Bottom: position angle of polarization, with the measurements of \citet{Bailey} and \citet{Lebofsky}, as above. }
\label{fig:polfig} 
\end{centering} 
\end{figure}

The
polarization, position angle and total flux spectra of NGC1068 
are shown in Figure~\ref{fig:polfig}, together with the polarimetric results of \citet{Bailey} and \citet{Lebofsky}.

%£££££££££££££££££££££££££££££££££££££££££££££££££££££££££££££££££££££

%££££££££££££££££££££££££££££££££££££££££££££££££££££££££££££££££££££

\section{Polarization Calculations}
\label{calcs}

To calculate the L-band polarization produced by coated dust grains, we have taken advantage of the discrete
dipole approximation (DDA) code, DDSCAT6.1 \citep{DDDA,Dflat,DF}. In
the DDA the dust grain is represented as an array of dipole
oscillators on a cubic lattice, and absorption, scattering and extinction
cross-sections are then calculated from the amplitudes of the dipoles
as they interact with the incident electric field.

The optical constants adopted for the grain cores and mantles are those
of  ``astronomical silicate'' \citep{DL84,D85,WD01} and the
organic refractory material of \citet{Li97}. The refractive index of
the organic component is derived from the Murchison meteorite and laboratory residues exposed to the solar UV field aboard the EURECA satellite \citep{Greenberg95,GL96}. While both of those materials show absorption bands at 5 - 9~$\mu$m that are not seen in the diffuse ISM \citep{PandA}, the 3.4~$\mu$m bands of both bear a good resemblance to the observed feature.

The existence of polarization caused by dust absorption indicates that
the dust grains must be aspherical, but beyond that their shapes are
not well-established. 
We
have treated the dust grains as oblate spheroids with an axial ratio
of 2:1, the cores and mantles having equal eccentricities. Based on fits to the Trapezium 9.7$\mu$m silicate feature,
\citet{DL84} find that 2:1 oblate spheroids are a reasonable
representation of dust grains, and \citet{Hil95} also favor oblate
over prolate spheroids.  They conclude that the axial ratio must be
$<$3:1, with a best-fitting value of 1.5:1. There has been some
dispute over whether oblate or prolate spheroids give a better fit to
the observational data \citep{GL96}, but we note that the main
discrepancy in the polarization predicted by these two shapes comes
over the 9.7$\mu$m silicate feature. The differences are small at the
wavelengths of interest in this work \citep[see Fig. 6
of][]{DL84}. The elongation of the grains will affect the efficiency
with which they polarize, but has little effect on the wavelength
dependence of the polarization \citep{GL96}.

We have adopted simple picket-fence alignment for
this work (which is equivalent to the perfect spinning alignment approximation for oblate grains), with the grains' short axes orientated perpendicular to the direction of propagation of the incident light. This has been shown to be quite adequate if only the wavelength dependence of
the polarization is required, rather than its absolute
value~\citep{CG90}.  The functional form used for the grain size distribution is the MRN $a^{-3.5}$ power law with limits of 0.005 and 0.25 $\mu$m \citep{MRN}. 

Given this information about the size, shape, composition and
orientation of the grains, DDSCAT calculates a number of
quantities. Those relevant to this work are 
\begin{eqnarray}
Q_{abs} & = & C_{abs}/\pi a^{2} \\
Q_{sca} & = & C_{sca}/\pi a^{2} \\
Q_{ext} & = & Q_{abs} + Q_{sca} \\
Q_{pol} & = & Q_{ext},\parallel - Q_{ext},\perp 
\end{eqnarray}
in which $Q$s are efficiencies and $C$s cross-sections for absorption,
scattering and extinction, and $Q_{ext},\parallel$ and $Q_{ext},\perp$
refer to extinction efficiencies for the two orthogonal incident
polarization states \citep{DDDA}.

From $Q_{ext}$ and $Q_{pol}$, extinction and polarization spectra for
ensembles of grains can be obtained in the following manner:
\begin{equation} \label{eq:tau}
\tau (\lambda)=N_{grain}
\int_{a_{min}}^{a_{max}}n(a)C_{geo}Q_{ext}(a,\lambda)da
\end{equation}
\begin{equation} \label{eq:pol}
p(\lambda)=N_{grain}\int_{a_{min}}^{a_{max}}n(a)C_{geo}\frac{Q_{pol}}{2}(a,\lambda)da
\end{equation}
where $C_{geo}$ is the geometrical cross-section of the grain in
question, $N_{grain}$ the column density of grains (with $n(a)$, the number of grains in the interval $a, a+da$, appropriately normalized) and $a_{min}$ and $a_{max}$ the limits of the size distribution
to be considered.

%££££££££££££££££££££££££££££££££££££££££££££££££££££££££££££££££££££

\section{Results and discussion}
\label{results}

We are interested in the magnitude of any increase in polarization
that would be expected over the 3.4 $\mu$m absorption band, given a
certain continuum polarization caused by dichroic absorption by
core-mantle grains. To obtain a crude estimate of the polarization that
might be expected over the 3.4 $\mu$m band, we can consider the polarization observed over the 9.7 $\mu$m silicate band in galactic lines of sight.  For a given grain shape, size and composition, changes in abundance and degree of alignment will alter the level of continuum polarization without affecting the ratio of feature to continuum polarization, $P_{feat}/P_{cont}$. In galactic lines of sight that can be fitted with pure absorptive polarization, this ratio is typically about 3-4, with a couple of objects reaching values of $\sim$7 (OMC1 BN, SgrA GCS3 II; \citet{Smith00}).  Alternatively, the optical constants of astronomical silicate give rise to $P_{feat}/P_{cont} \approx 12$ \citep{DL84,Hil95}.
The "excess" polarization over an absorption feature is determined by the strength of the band, and \citet{LG02} have shown that silicates and hydrocarbons polarize to a similar degree per unit optical depth. Observed values of $\tau_{9.7}/\tau_{3.4}$ range from about 13-19 \citep{S95}, implying  $P_{feat}/P_{cont}$ $\sim$1.15 - 2.0 in the 3.4 $\mu$m band. For NGC1068, where the L-band continuum polarization is about 2.5\%, this translates to peak polarizations of 2.9 - 5\% across the 3.4 $\mu$m feature. This suggests that a significant excess polarization should be detectable over the 3.4 $\mu$m feature, although the spread in silicate feature polarizations and $\tau_{9.7}/\tau_{3.4}$, and the fact that most of the sources in \citet{Smith00} are dominated by molecular cloud material mean that these numbers should be treated with some caution.

\begin{figure} [t]
\begin{centering}
\includegraphics[width=90mm,angle=0]{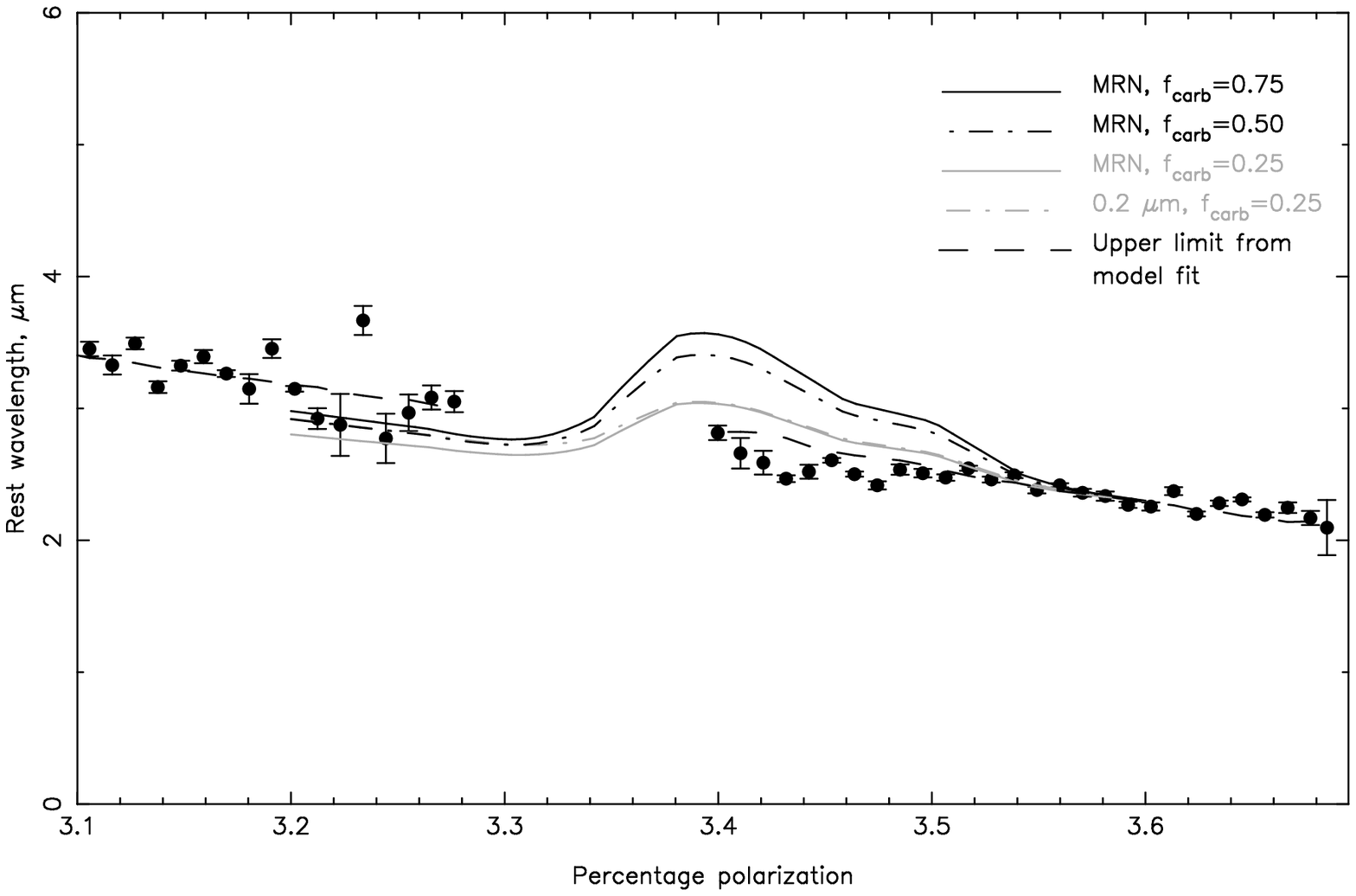}
%from /Users/rmason/data/1068_Lpol/new_May2006/data+reduction/finishpol.f90
\caption{\small{Observed and predicted polarization over the 3.4 $\mu$m feature in NGC1068 for an MRN distribution of coated silicate particles with different thicknesses of mantle. Also shown are the polarization of a 0.2 $\mu$m grain ($f_{carb}$= 0.25), close to the largest size in the MRN distribution, and the limit on the change in polarization over the feature based on a model fit to the observed polarization (see text). The column density of grains (eq. 6) was chosen to reproduce the observed polarization around 3.55 $\mu$m, a region expected to be unaffected by the 3.4~$\mu$m feature.}}
\label{MRN} 
\end{centering} 
\end{figure}

To estimate the likely maximum observed polarization change over the 3.4~$\mu$m band in NGC~1068,  the best-fitting linear continuum was first found for the polarization data. A model polarization spectrum was then created from the optical depth spectrum derived from the spectropolarimetry data, assuming that $\rm p \propto \tau$. The model spectrum was normalized to the continuum around 3.55~$\mu$m, then scaled by various factors and the reduced $\chi^{2}$ between it and the binned polarization spectrum calculated for each scaling. The best fit corresponds to a deviation of -0.12$\pm$0.25\% (95\% confidence) from the linear continuum; the data are in fact consistent with a small {\em decrease} in polarization over the 3.4~$\mu$m feature. The model fit giving the upper limit on the change in polarization over the 3.4~$\mu$m band (i.e., peaking at 0.13\% above the continuum) is shown in Figure~\ref{MRN}.

The polarization calculated as described in \S~\ref{calcs} for a population of core-mantle grains
with the MRN size distribution is also shown in Fig. \ref{MRN}. As grain size may affect the magnitude of the 3.4 $\mu$m polarization excess (\S\ref{Var}), the polarization from a 0.2 $\mu$m grain, close to the largest size in the MRN distribution and somewhat larger than the  average $<a> = 0.087 \mu$m of the finite cylinders in the \citet{Li97} model, is also shown.
The different curves in the figure result from different values of $f_{carb}$, the fraction of the grain volume that is contained in the mantle. Various figures have been suggested for this quantity;  \citet{LG02} consider $f_{carb}=0.2,0.5,0.75$ while according to \cite{CG90}, $f_{carb}\sim0.9$.  \citet{GL96} and \citet{Li97} estimate mantle/core volume ratios of about 2:1.
 As expected, the polarization at the peak of the 3.4 $\mu$m band increases with the amount of carbonaceous material in the grains, from approximately 3.0\% for the 25\% -mantle MRN grains, to about 3.6\% for the 75\%-mantle particles. 
 
For the core/mantle ratios closest to those most commonly quoted in the literature, the 3.4~$\mu$m polarization predicted by the model is clearly well in excess of the observed polarization. In the case of the grains with only 25\% of the volume contained in the mantle, the polarization at the peak of the 3.4~$\mu$m absorption is 0.29\% over the linear continuum fit mentioned above. This is still somewhat above the limit we derive on the observed polarization; the expected and observed polarizations are not  reconciled even with only a small fraction of each grain being composed of organic refractory material. The core-mantle grain model, at least in the form proposed for the Galactic diffuse ISM,  is unlikely to explain the L-band  polarization of NGC~1068.

\subsection{Variations on the basic core-mantle model}
\label{Var}

Are there likely to be significant differences in the grain population(s) in NGC1068 that might diminish the amount of polarization expected over the 3.4 $\mu$m band, even while core-mantle grains are present? The similarity of the band profile in NGC 1068 and the Galactic diffuse ISM suggests that the composition of the hydrocarbon material is similar in both galaxies \citep{Dogtanian, Mason04}, but other differences may arise. For instance, as the \citet{Li97} model proposes that the entire infrared polarization arises in the core-mantle grains, we have so far also assumed this, but if another, non-hydrocarbon-containing aligned grain component also contributed to the polarization in NGC1068 it would diminish the size of the rise in polarization expected over the 3.4 $\mu$m band while contributing to $A_{V}$ but not $\tau_{3.4}$.  In NGC1068, such a second polarizing grain population could conceivably be, for example, bare silicates arising from destruction of the mantles on some fraction of the grains.

Values of extinction to the complex, infrared-emitting regions of the nucleus of NGC1068 may not be as straightforward to interpret as the extinction through the large column of cold dust towards the Galactic Center, but estimates range from $\sim$15 \citep{Lumsden99,Watanabe} to $\sim$40 \citep{Young95}, disregarding values based on $\tau_{3.4}$. Given that $\tau_{3.4}\approx$ 0.1, this implies $A_{V}/\tau_{3.4}\sim150-400$, compared with galactic values of $\sim$150 \citep[towards the galactic center;][]{P94} or  $\sim$several hundred \citep[towards field stars at various galactic longitudes;][]{Rawlings03}. There is therefore no compelling reason to think that grain components other than those proposed for the galactic diffuse ISM must exist in NGC1068, but the wide range of estimates of $A_{V}/\tau_{3.4}$ for both the galactic and extragalactic lines of sight means that this cannot be ruled out either. 

Another issue that could affect the 3.4~$\mu$m band in both extinction and polarization is that of grain size. Increasing grain size tends to increase the efficiency of continuum extinction 
while decreasing the strength of absorption features, so large grains might be able to polarize the L-band continuum effectively without producing much excess polarization through the 3.4 $\mu$m band.
For grains much larger than about 0.2 $\mu$m in radius, scattering starts to become important ($Q_{abs}/Q_{sca}\approx5$ for a 0.2 $\mu$m core-mantle 2:1 oblate spheroid at 3.3 $\mu$m) and calculations of the polarization from such particles in a complex system like NGC1068, where the inclination of the dusty torus and multiple scattering effects may be critical, are beyond the scope of this paper.  

It has been suggested that dust grains in AGN may be biased towards larger sizes than in the diffuse ISM, but the evidence remains contradictory.
 \citet{Laor93} pointed out that large
($\sim$10 $\mu$m) dust grains are likely to
survive longer than small grains close to an AGN and will not produce a silicate emission feature. The weakness or absence of silicate emission in type 1 AGN prompted the inclusion of large grains in some torus models \citep{vanBem}, but its recent discovery in several quasars  may argue against large grains \citep{Hao05,Siebenmorgen05}. 
Flat L-M$^{\prime}$ colors, peculiar $E_{B-V}/N_{H}$ and
$A_{V}/N_{H}$ ratios in Seyfert 2 nuclei and the absence of the
2175\AA~feature in reddened Seyfert 1s have all been interpreted as evidence for large grains \citep{Imcols,MaioII,MaioI}, but geometrical effects may also be able to explain many of these observations \citep{Wein}.  In the specific case of NGC1068, both \citet{Young95} and \citet{Watanabe} were able to fit the observed J- to K-band  and optical polarization with grain sizes no different from those thought to exist in the galactic DISM, although Watanabe et al.  suggest that scattering from large grains in the torus might be a viable alternative. 
Detailed extinction and polarization calculations such as those of \citet{Watanabe}, if extended to the 3.4 $\mu$m band, could provide valuable constraints on both the size and structure of the grains in AGN tori.

Finally, we note that if the grain size distribution is biased towards larger sizes in NGC1068, this could be through preferential destruction of small grains, or large grains may have grown by coagulation in this warm, dense environment. If the latter, then they may have lost some of their former core-mantle nature. It seems likely that small hydrocarbon inclusions in large, coagulated grains would leave their polarization signature on the 3.4 $\mu$m band, but again, further calculations would be needed to test this.

\subsection{Hydrocarbon formation in the diffuse ISM}
\label{Alt}

The absence of excess polarization over the 3.4 $\mu$m band is naturally explained if the feature arises in a population of grains that does not contribute significantly to the continuum polarization. Such a grain population must be small and/or optically isotropic \citep[small grains being harder to align than large ones;][and references therein]{Laza}, and have no physical connection to the silicate grains. Recent laboratory work has provided persuasive evidence that small carbon grains like those thought to be ejected from AGB stars can be hydrogenated during the later stages of stellar evolution \citep{Schnaiter} and in the diffuse ISM \citep{Me99,Munoz01,Me02}.  Calculations indicate that hydrogenation proceeds at a fast enough rate in the diffuse ISM to balance the dehydrogenation caused by  UV photons. Conversely, in dense molecular clouds, the reduced abundance of atomic H and the presence of icy mantles act to prevent rehydrogenation of the grains, which can still be efficiently dehydrogenated by photons and cosmic rays \citep{Me01,Mennella03}. These and other lines of evidence \citep[e.g.][]{Shenoy} imply that most hydrocarbon-containing grains are formed in the diffuse ISM and that re-formation dominates over the dehydrogenation that subsequently occurs, in agreement with the non-detection of the 3.4 $\mu$m feature in dense cloud material. This evolutionary scenario for the aliphatic hydrocarbons is entirely consistent with the lack of a 3.4~$\mu$m polarization excess in NGC~1068 and several Galactic lines of sight \citep{A99, Ishii, Chiar06}.

\section{Summary}
\label{conc}

We have performed L-band spectropolarimetry of NGC~1068 and shown that the excess polarization over the 3.4~$\mu$m feature is below that which would be expected on the basis of the silicate-core/organic-mantle grain model as applied to the galactic diffuse ISM, consistent with a growing body of evidence suggesting that the aliphatic hydrocarbons in the general diffuse ISM are not formed by processing of the ice mantles that form on silicate grains in molecular clouds. The coated grain model could still be valid in NGC~1068 if there also exists an extra, non-hydrocarbon aligned grain population, or possibly if the grain size distribution is biased to larger sizes than in the diffuse ISM of our galaxy (detailed calculations of both the continuum and 3.4~$\mu$m feature polarization from micron-sized dust grains may be a useful way of constraining the size and/or composition of the carbonaceous grain population in AGN).
Alternatively, reaction of small carbon grains with atomic hydrogen in the diffuse ISM would be expected to produce a population of small, nonpolarizing hydrocarbon-containing grains which would naturally explain the lack of polarization of the 3.4~$\mu$m feature.  Such a model is also successful in accounting for the non-detection of the 3.4~$\mu$m band in molecular clouds, which is otherwise difficult to explain.  

\section{Acknowledgments}

We would like to thank R. Antonucci,  P. Hirst, M. Kishimoto,  T. Roush, M. Smith and A. Tielens for taking the time to  comment, and A. Li for providing the optical constants  in
tabular form. We are grateful to the anonymous referee for comments that strengthened the paper. We thank the Department of Physical Sciences, University of Hertfordshire for providing IRPOL2 for the UKIRT. The United Kingdom Infrared Telescope is operated by the Joint Astronomy center on behalf of the U.K. Particle Physics and Astronomy Research Council. Supported by the Gemini Observatory, which is operated by the Association of Universities 
for Research in Astronomy, Inc., on behalf of the international Gemini partnership of Argentina, 
Australia, Brazil, Canada, Chile, the United Kingdom, and the United States of America.

\end{document}